\documentclass[aip,pop,showpacs,numerical,reprint,floatfix]{revtex4-1}
\usepackage{dcolumn}
\usepackage{bm}
\usepackage{amsopn,amsmath,amssymb,amsfonts}
\usepackage{color}
\usepackage{graphicx}

\begin{document}
\title{Nonlinear structures and anomalous transport in partially magnetized
$\mathbf{\ E\times B} $  plasmas.}
\author{Salomon Janhunen}
\email{e-mail: salomon.janhunen@usask.ca.}
\affiliation{University of Saskatchewan, 116 Science Place, Saskatoon, SK S7N 5E2
Canada}
\author{Andrei Smolyakov}
\affiliation{University of Saskatchewan, 116 Science Place, Saskatoon, SK S7N 5E2
Canada}
\author{Oleksandr Chapurin}
\affiliation{University of Saskatchewan, 116 Science Place, Saskatoon, SK S7N 5E2
Canada}
\author{Dmytro Sydorenko}
\affiliation{University of Alberta, 3-235 Centennial Centre for Interdisciplinary Science
Edmonton, AB T6G2E9, Canada}
\author{Igor Kaganovich}
\affiliation{Princeton University / Princeton Plasma Physics Lab, 100 Stellarator Rd, Princeton 08543-0451, USA}
\author{Yevgeni Raitses}
\affiliation{Princeton University / Princeton Plasma Physics Lab, 100 Stellarator Rd, Princeton 08543-0451, USA}
\date{\today }
% \pacs{Valid PACS appear here}% PACS, the Physics and Astronomy
%                              % Classification Scheme.
% \keywords{Suggested keywords}%Use showkeys class option if keyword
%                               %display desired

\begin{abstract}
Nonlinear dynamics of the electron-cyclotron instability driven by the
electron $\mathbf{\ E\times B}$ current in crossed electric and magnetic field is studied. In nonlinear regime the instability proceeds by developing a large amplitude coherent wave driven by the energy input from the fundamental cyclotron resonance. Further evolution shows the formation of the long wavelength envelope akin to the modulational instability. Simultaneously, the ion density shows the development of high-k content responsible for wave focusing and sharp peaks on the periodic cnoidal wave structure.  It is shown that the anomalous electron transport (along the direction of the applied electric field) is dominated by the long wavelength part of the turbulent spectrum.
% Valid PACS numbers may be entered using the \verb+\pacs{#1}+ command.
\end{abstract}

\maketitle

% Frame for the paper:
%
%
%
%  Electron cyclotron drift instability (ECDI)
%
%  ECDI very effective in mixing phase space, more so than two-stream
%
%  Even in the linear stage ECDI appears to be affected by non-linear interactions,
%  appearing as preference for low-k mode growth over higher-k modes.
%
%  In the non-linear stage, mode structure peaks on low-k modes
%  (lower than the lowest EC resonance), which are most effective in driving anomalous
%  current.
%
%  Electron density structures are mostly low-k, but ion density has a cnoidal-like
%  structure to it, which explains the equal spacing of peaks in the \omega and k spectra.
%

\section{Introduction}
Partially magnetized weakly collisional plasmas with magnetized
electrons and weakly magnetized ions are abundant in nature and
laboratory conditions. Therefore, their nonlinear behavior is of
considerable interest for fundamental physics and applications. One of
the most common examples is a plasma discharge driven by transverse
current perpendicular to the magnetic field,
\cite{Buneman1962,ArefevNF1969,WongPF1970,GaryJPP1970} either due to
free streaming of unmagnetized ions across the magnetic field or due
to the electron drift current in crossed electric and magnetic field,
$\mathbf{V}_{E}=\mathbf{\ E\times B/}B^{2}$. Such configurations are
relevant to collisionless shock waves in space, pulsed power
laboratory devices, Penning discharges and various devices for
material processing and space propulsion. 
Plasmas with crossed $\mathbf{E}\times\mathbf{B}$ fields are subject to a variety of instabilities such as ion-sound, lower-hybrid and Simon-Hoh
modes which may be driven by plasma density, magnetic field and temperature gradients as well as collisions \cite{LashmoreNF1973,SmolyakovPPCF2017,DavidsonNF1975}. 
The electron cyclotron drift instability is of particular interest
because it does not require any gradients and may be active in a homogeneous collisionless plasma with electric field perpendicular to the magnetic
field.\cite{SizonenkoNF1967,GaryJPP1970,WongPF1970}
Large-amplitude waves present in satellite observations of bow shock
crossings have been associated with current driven electron-cyclotron instabilities.\cite{WilsonJGR2010,WilsonJGR2014-1,WilsonJGR2014-2,BrenemanJGR2013}
Presence of the electron cyclotron instabilities has been confirmed by numerical simulations of bow shocks. \cite{MuschiettiAdvSpR2006,MuschiettiJGR2013,MatsukiyoJGR2009}

There have been a number of earlier studies
\cite{BiskampPRL1971,BiskampNF1972,BiskampPF1973,ForslundPRL1970,LampePRL1971,LampePRL1971,LominadzeJETP1973}
addressing linear and nonlinear theory of the electron-cyclotron instabilities, but many critical questions remained unresolved. Recent
developments in applications of $\mathbf{E\times B}$ discharges (also
referred as $\mathbf{ E\times B}$ plasma below) such as HiPIMPS
magnetrons, Hall thrusters and Penning discharges have again raised
questions on the nature of turbulence, transport and nonlinear
structures in such conditions
\cite{TsikataPRL2015,SmirnovIEEE2004,AdamPoP2004,BoeufPRL2013,BoeufFP2014,BoeufJAP2017}.

Linearly, the electron-cyclotron  instability is based on
the interaction of the electron cyclotron mode with ion plasma
oscillations. Both dissipative and reactive regimes may occur. In the dissipative regime, the  negative energy wave is excited due to resonance absorption of wave energy by  electron and ions.\cite{SizonenkoNF1967} The  reactive instability may occur  due to coupling of waves with positive and negative energy.\cite{LashmorePhysA1970,BiskampNF1972} For
propagation strictly perpendicular to the magnetic field and electrons
subject to $\mathbf{E\times B}$ drift, the resonant condition is
$\omega -\mathbf{k\cdot v}_{E}-m\Omega _{ce}=0$. It has been noted
that electron cyclotron drift instability (ECDI) due to linear and/or nonlinear effects
\cite{SizonenkoNF1967,ArefevTechPhys1970,LashmoreNF1973,LampePRL1971,LampePF1972a} may in some regimes become similar to the ion-sound instability
in unmagnetized plasmas. The transition of the ECDI instability, which in
essential way depends on the presence of the magnetic field, into the regime
which resembles the ion sound instability in absence of the magnetic field
has become a common theme of many earlier works in the literature
\cite{LampePF1972b,LampePF1972a}. In recent years, the regime of
unmagnetized ion sound turbulence has been considered as a main
paradigm for nonlinear regime of the electron cyclotron drift
instability --- in particular --- for calculations of the associated
anomalous current in Hall thrusters \cite{KatzIEPC2015,CavalierPoP2013,LafleurPoP2016a,LafleurPoP2016b}.

The goal of this paper is to investigate the nonlinear regime of the
ECDI instability, its possible transition to the unmagnetized
ion-sound regime, and associated level of the anomalous transport. We
show here that for typical plasma parameters relevant to
applications to magnetron and Hall thruster plasmas, the ion-sound
like regime of the ECDI (with fully demagnetized electrons) does not
occur, even in absence of energy losses for electrons. The magnetic
field continues to play an important role in the electron dynamics,
particularly in the energy supply to the mode and electron heating mechanism.  Nonlinearly the instability continues to exist as coherent
mode at the fundamental cyclotron resonance $k_{0}\equiv
v_{E}/\Omega_{ce}$. Interestingly, electron demagnetization during
ECDI has recently been discussed in applications to the collisionless
bow shock plasma of the Earth.\cite{MuschiettiJGR2013} Nonlinear
simulations of Ref.~\onlinecite{MuschiettiJGR2013} have also shown
that the fundamental cyclotron resonance remains active and no full
demagnetization occurs.

Our results also demonstrate that the injected energy (primarily at the
lowest resonance) cascades toward even longer wavelength modes. This
inverse energy cascade (toward longer wavelengths) is characterized by
the formation of a long wavelength envelope, similar to the
modulational instability of the wave packets. We posit that the slow
long wavelength envelope discovered in our simulations is responsible
for low frequency structures exhibited by ECDI instability recently
observed experimentally in high-power pulsed magnetron (HIPIMS)
discharge.\cite{TsikataPRL2015}  We investigate the anomalous current
and find that it is dominated by the long-wavelength modes.

\section{Linear dispersion relation}

In this section  we discuss main features of the electron-cyclotron drift
instability (ECDI) with respect to the linear dispersion relation that was obtained in a number of earlier works.\cite{GaryJPP1970,WongPF1970,ArefevTechPhys1970} 
We consider a plasma immersed in the crossed electric and magnetic field, $\mathbf{E}=E_{0}\widehat{\mathbf{z}}$, $\mathbf{B}_{0}=B_{0}\widehat{\mathbf{y}}$. The
ions are unmagnetized, but electrons are magnetized and experience the
$\mathbf{E\times B}$ drift, $\mathbf{v}_{E}=-E_{0}/B_{0}\widehat{\mathbf{x}}$.
One-dimensional fluctuations are propagating in the $x$ direction,
$\mathbf{k}=k\widehat{\mathbf{x}}$. The linear kinetic dispersion relation
\cite{GaryJPP1970,WongPF1970,ArefevTechPhys1970} has the form
$1+K_{i}+K_{e}=0$, where the ion response is $K_{i}=-1/\left( 2k^{2}\lambda
_{Di}^{2}\right) Z^{^{\prime }}\left( \omega /\sqrt{2}kv_{i}\right) $, and
the electrons are described by
\begin{eqnarray}
&&K_{e}=\frac{1}{k^{2}\lambda _{De}^{2}}\left[ 1-\exp \left( -k^{2}\rho
_{e}^{2}\right) I_{0}\left( k_{\bot }^{2}\rho _{e}^{2}\right) \right.  \notag
\label{eq:dispersion-magn} \\
&&\left. -2\left( \omega -k_{x}v_{E}\right) ^{2}\sum\limits_{m=1}^{\infty}%
\frac{\exp \left( -k_{\bot }^{2}\rho _{e}^{2}\right) I_{m}\left( k_{\bot
}^{2}\rho _{e}^{2}\right) }{\left( \omega -k_{x}v_{E}\right)
^{2}-m^{2}\Omega _{ce}^{2}}\right]
\end{eqnarray}%
where $\Omega _{ce}=eB/m_{e}$, $\lambda _{D\alpha }=\varepsilon
_{0}T_{\alpha }/\left( e^{2}n_{0}\right) $, $v_{\alpha }=\sqrt{T_{a}/m_{a}}$
for species $\alpha $, $I_{m}$ is the modified Bessel function of the first
kind, and $Z(z)$ is the plasma dispersion function.

Unstable eigenmodes
form a discrete set of modes localized near the resonances $\omega
-kv_{E}=m\Omega _{ce}$. In the cold plasma limit, only the lowest $m=1$ resonance exists and Eq.~(\ref{eq:dispersion-magn}) reduces to the reactive Buneman
instability \cite{Buneman1962} with the dispersion relation $1=\omega
_{pi}^{2}/\omega ^{2}+\omega _{pe}^{2}/\left( \left( \omega -kv_{E}\right)
^{2}-\Omega _{ce}^{2}\right)$, which was discussed as a mechanism of anomalous 
transport in $\mathbf{E\times B}$ discharges in Refs.~\onlinecite{BaranovAIAA1996,
AdamPoP2004}. The finite electron temperature makes $m=1$ resonance narrow
and opens up higher $m$ resonances at $kv_{E}\simeq m\Omega_{ce}$. The
growth rates of the higher resonance modes are first increasing with $m$
and then decrease for high $m$. The width of the resonances decreases with
temperature \cite{LominadzeJETP1973,LashmorePhysA1970}. 
A detailed structure and behavior of the linear eigenmodes from Eq.~(\ref{eq:dispersion-magn}) was investigated in a number of papers, such as in Refs.~\onlinecite{GaryJPP1970,LominadzeJETP1973,AdamPoP2004,DucrocqPoP2006,CavalierPoP2013}.
A recent discussion of properties of the linear ECDI instability can also be
found in Ref.~\onlinecite{MuschiettiJGR2013}.  

\section{Nonlinear dynamics and formation of the long-wavelength envelope}

Nonlinear dynamics of the ECDI instability is studied here with 1D3V
parallel particle-in-cell simulations using the PIC code EDIPIC
\cite{SydorenkoTh2006}.  As a characteristic example we consider a
xenon plasma ($m_{Xe}=131.293\,\text{amu}$) with Hall-effect thruster
relevant parameters of $n_{0}=10^{17}\,\text{m}^{-3}$,
$E_{0}=20\,\text{kV/m}$, $B_{0}=0.02\,\text{T}$, initial
temperatures of $T_{e}=10\,\text{eV}$ and $T_{i}=0.2\,\text{eV}$,
simulation box length $L=44.56\,\text{mm}$ using a spatial resolution
in $x$ of $\lambda_{De}/8$, and the initial electron Larmor radius of
$0.5\,\text{mm}$. The wave vector is constrained by periodicity of the
simulation domain to $k\,L=2\pi\,n$, where $L=2\pi r$ is the azimuthal
length of the channel (or periodic portion thereof). The particles are
initialized as Maxwell-Boltzmann distributions, shifted by the
$\mathbf{E}\times \mathbf{B}$ drift velocity $v_{E}$ for the
electrons. Time step is chosen to fulfill the CFL condition for
particles up to $35v_{e}$ from the initial value, and $10^{4}$ marker
particles per cell are used for a noise level of 1\% or less.

\begin{figure}[thp]
  \includegraphics[width=0.975\columnwidth,viewport=52 61 783 585,clip]{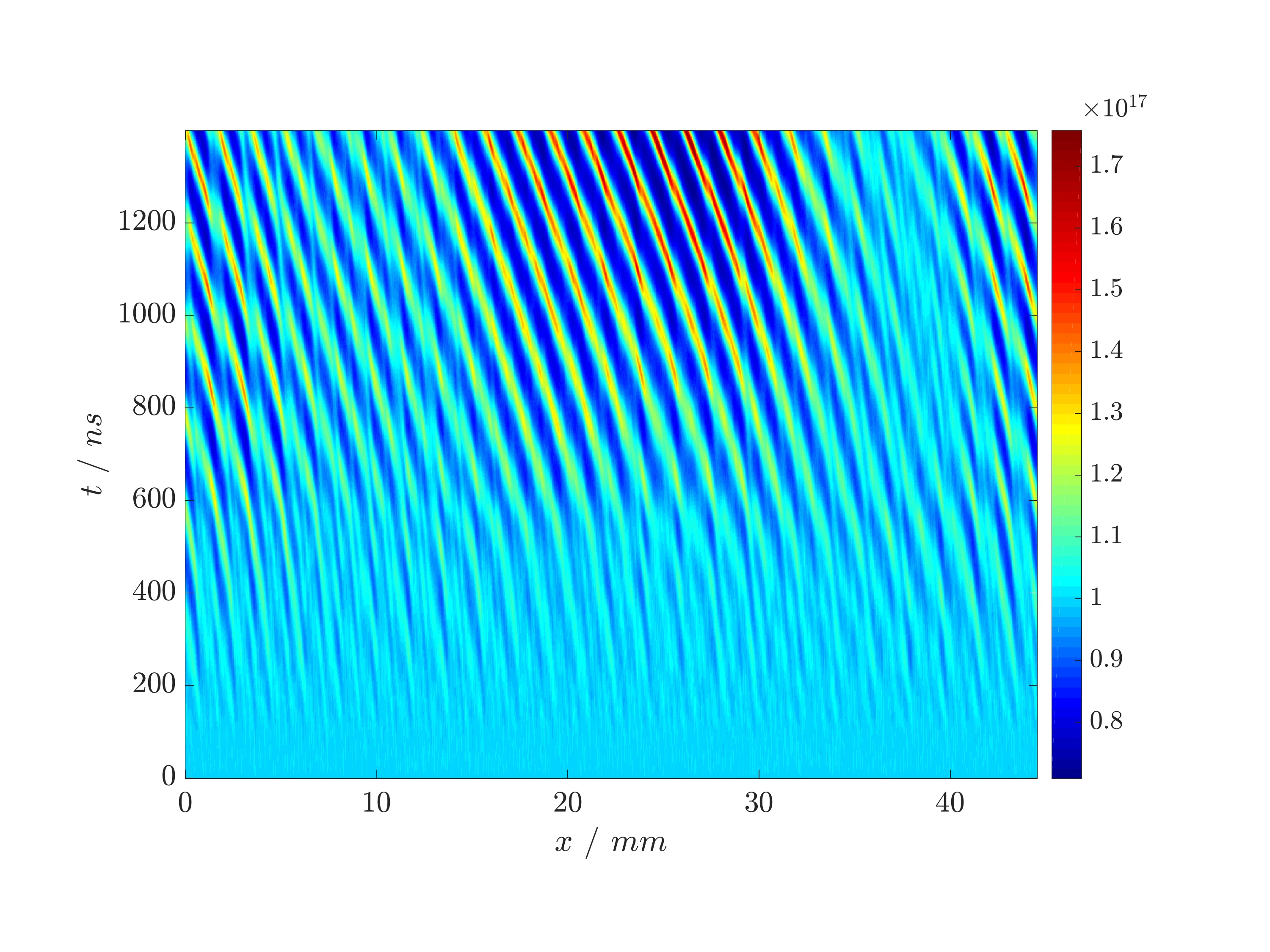}
  \caption{Ion density as a function of time. Note the simultaneous appearance of the cnoidal structure along with modulation.}
  \label{fig:density-b}
\end{figure}

\begin{figure}[thp]
  \includegraphics[width=0.975\columnwidth,viewport=67 73 744 551,clip]{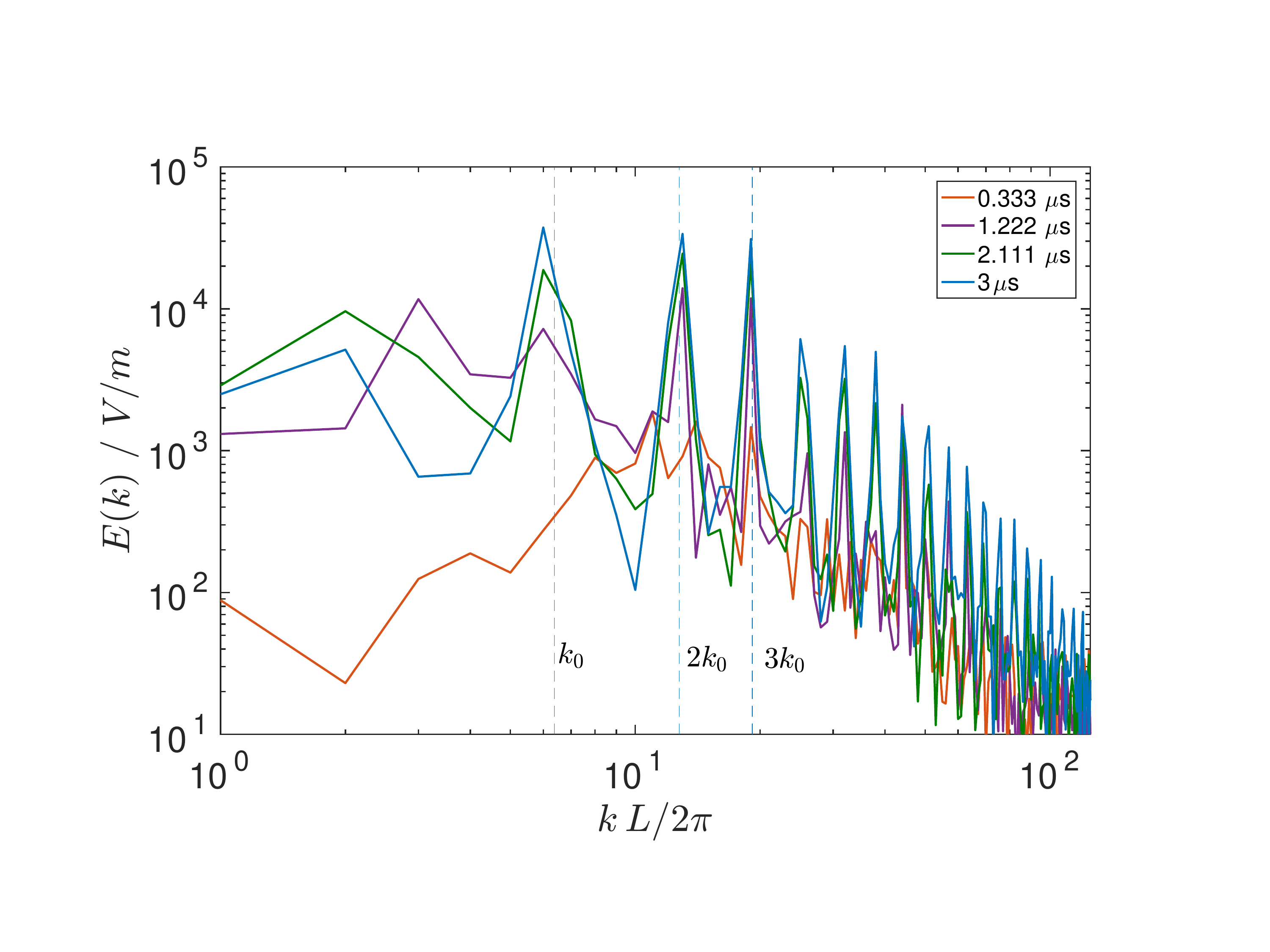}
  \caption{Amplitudes of $E_{x}$ spectral components over the simulation time. First three cyclotron harmonics are shown as vertical lines. Significant sub-resonant components are seen below the lowest cyclotron harmonic, and an upward cascade to lower $k$.}
  \label{fig:density-c}
\end{figure}

The linear instability commences with the growth of the most unstable linear
cyclotron harmonic (for our parameters here $m=3$ and $n=6$). At a later
time, the progressively lower $k$ cyclotron harmonics take over as
Figs.~\ref{fig:density-b} and \ref{fig:density-c} illustrate. 
  
\begin{figure}[thp]
  \includegraphics[width=0.975\columnwidth,viewport=35 47 730 552,clip]{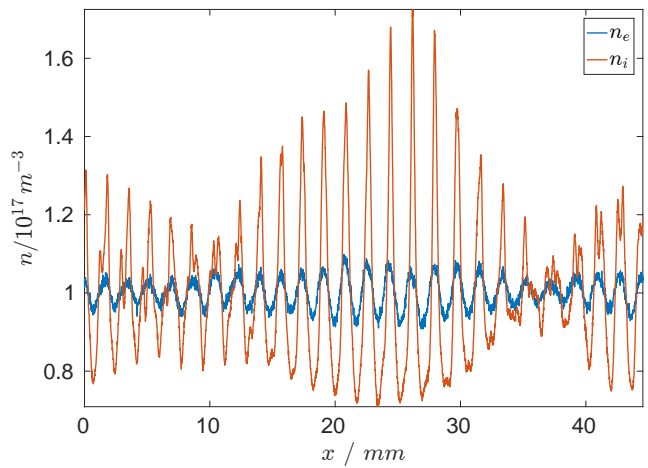}
  \caption{Ion and electron density shown at $1.4\,{\protect\mu }\text{s}$. Note the cnoidal structure of ion density fluctuations.
}
\label{fig:density-a}  
\end{figure}

\begin{figure}[thp]
  \includegraphics[width=0.975\columnwidth,viewport=84 51 769 556,clip]{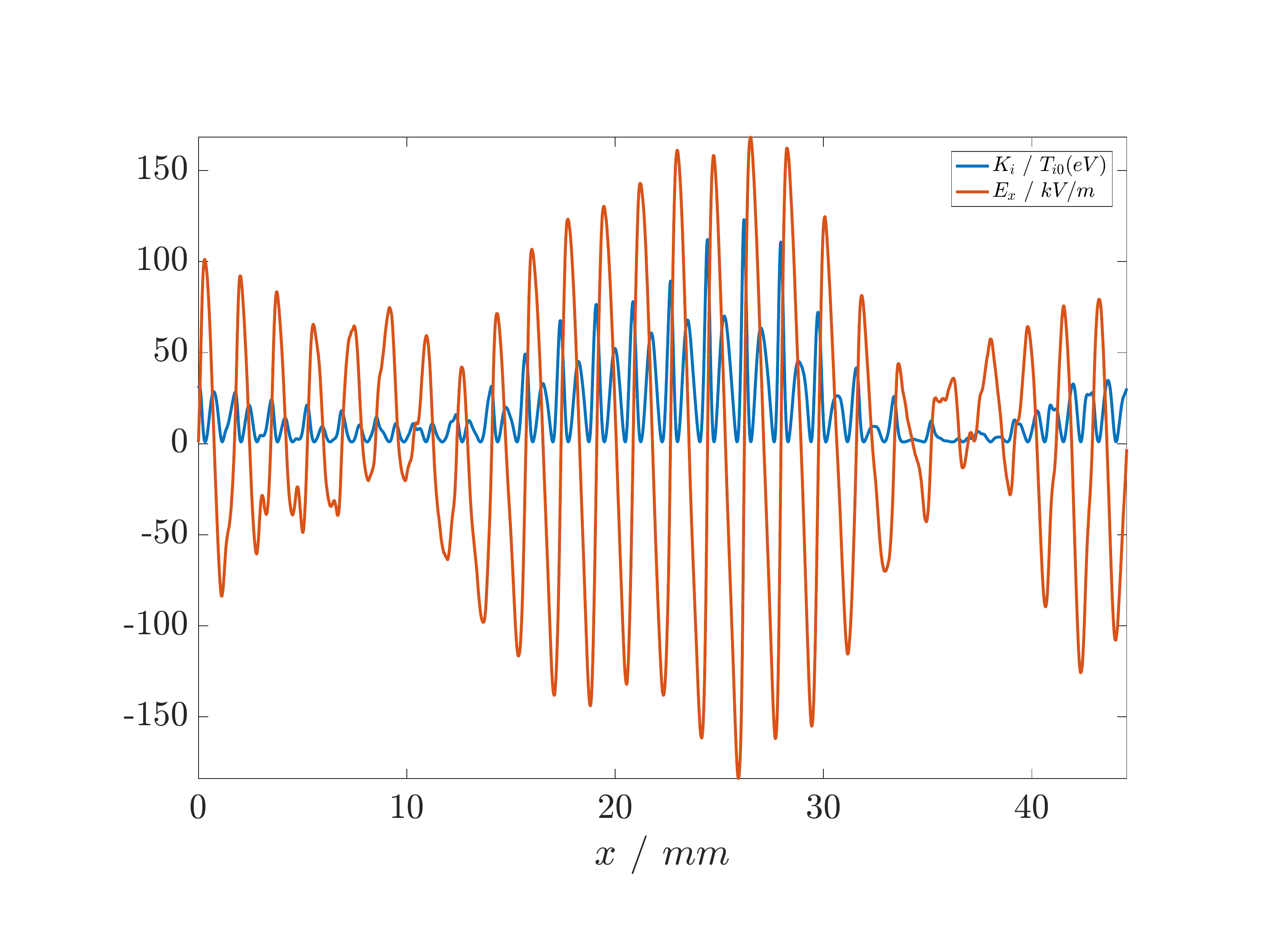}
  \caption{Electric field and ion temperature at $1.4\,{\mu }\text{s}$.}
  \label{fig:density-d}
\end{figure}

In part, the downward shift occurs due to increase of the electron
temperature as a result of the heating \cite{MuschiettiJGR2013}.  In
nonlinear stage however this tendency is amplified by the inverse
cascade which shifts energy further down to large scales much below of
the length scale of the fundamental cyclotron mode
$k_{0}^{-1},k_{0}=\Omega _{ce}/v_{E}$, as evidenced in
Fig.~\ref{fig:density-c} as well as by the modulation of the wave
envelope in Figs~\ref{fig:density-a} and \ref{fig:density-d}. Note
that in our simulations modes corresponding to the few lowest cyclotron
harmonics with $m<10$ remain to be clearly present well into the nonlinear
stage as seen in Fig.~\ref{fig:density-c}.

\begin{figure}[thp]
  \includegraphics[width=0.975\columnwidth,viewport=51 66 732 561,clip]{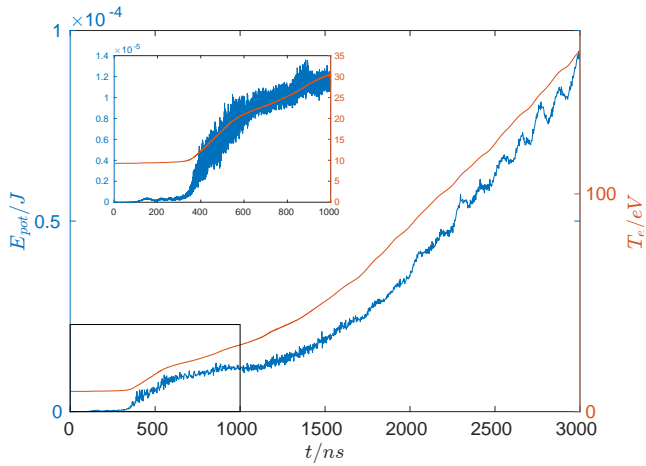}
\caption{Potential energy and electron temperature in the simulation.}
\label{fig:enetempdist-a}
\end{figure}

\begin{figure}[thp]
  \includegraphics[width=0.975\columnwidth,viewport=95 50 758 591,clip]{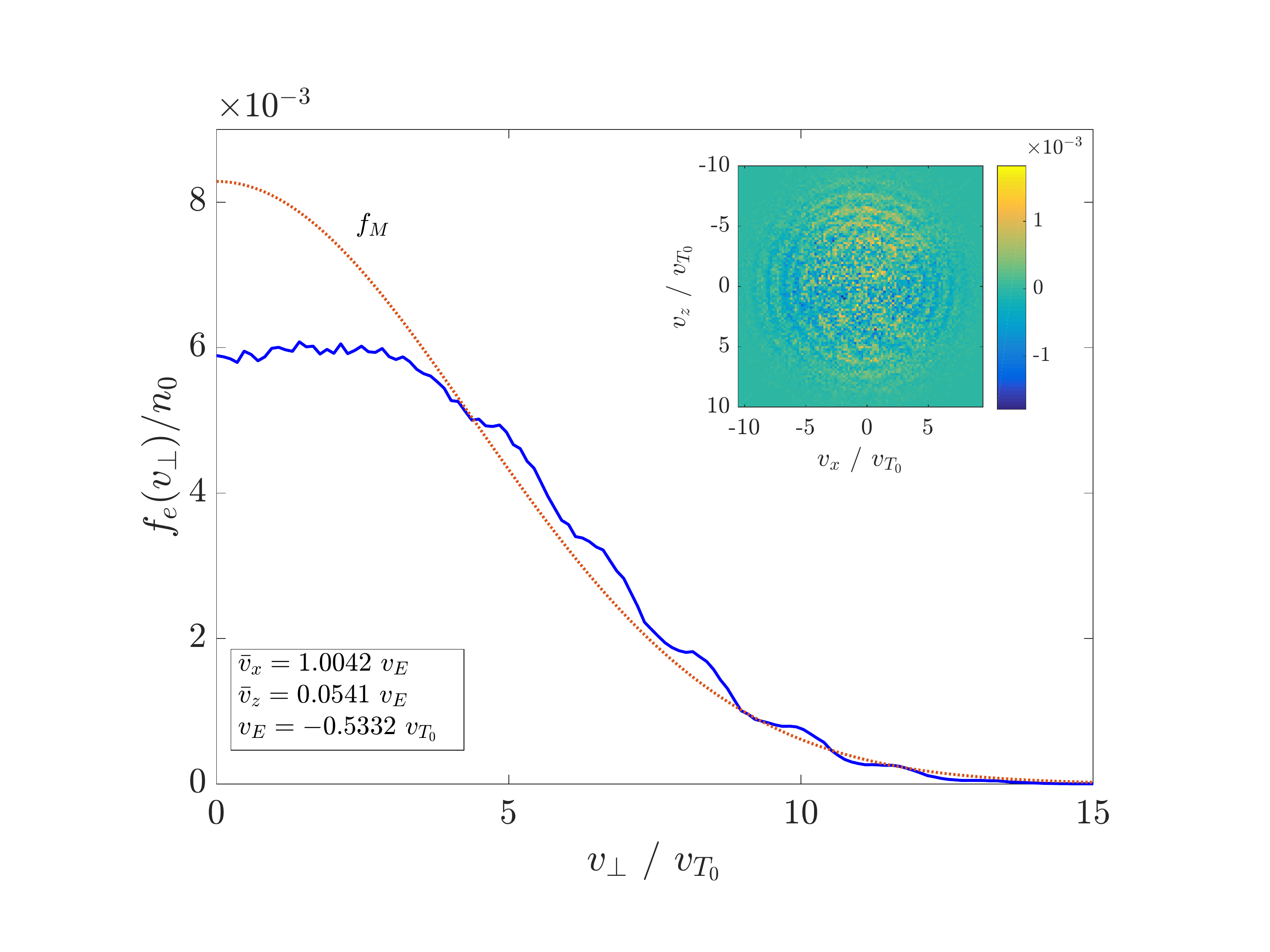}
  \caption{Distribution function of electrons at $t=1.4\protect\mu{s}$, with annular patterns apparent due to heating. The gyro-angle dependent part is shown in
    the insert. Here $v_x$ is the velocity in $E\times B$ direction (in units of the $v_E$), $v_z$ is in the direction of the background electric field, and $v_{\text{th}}$ is electron thermal velocity. The Maxwellian distribution ($f_M$) given as reference has the same temperature and volume as the measured one.}
  \label{fig:enetempdist-b}
\end{figure}

The $\mathbf{E\times B}$ instability described above is a very effective
mechanism for electron heating due local trapping and detrapping in the
time dependent potential formed by the magnetic field and the wave field
.\cite{BiskampNF1972,ForslundPF1972a,ForslundPRL1970} Even when
initiated with almost cold electrons of $T_{e}=0.001\,\text{eV}$, the
electron temperature rises to $T_{e}=20$ eV within few
$\gamma^{-1}\simeq \omega _{LH}^{-1}$. Within this time range, the
instability changes from the linear exponential growth to the slower
growth in which the potential energy and electron temperature increase
at the same rate approximately linearly in time, as shown in
Fig.~\ref{fig:enetempdist-a}. The electron heating is manifested as
intense phase-space mixing of the electron distribution function which
becomes flattened.\cite{DucrocqPoP2006} The flattening of the
distribution can be visualized through the {\it excess kurtosis} of
the distribution function, defined as $\text{Kurt}(f)=\int{(x-\mu)^4
  f(x)/\sigma^{4}}\,dx-3$, with mean $\mu$ and variance $\sigma^2$. It
becomes apparent from Fig.~\ref{fig:enetempdist-b} that heating has
flattened the distribution away from the Maxwell-Boltzmann statistics,
and caused a slight asymmetry in the $x-z$ temperatures. The
development of finite excess kurtosis and change of the distribution
from Maxwellian (with kurtosis 3) to platykurtic (kurtosis of
less than 3), which occur at $t\approx 100\,\text{ns}$,
Fig.~\ref{fig:density-d}, mark the transition from the linear
exponential growth to the slower nonlinear regime. The flattened
distribution is also observed in the ionosphere, like in
Ref.~\onlinecite{MozerJGR2013}.

\begin{figure}[thp]
    \includegraphics[width=0.975\columnwidth,viewport=68 93 759 546,clip]{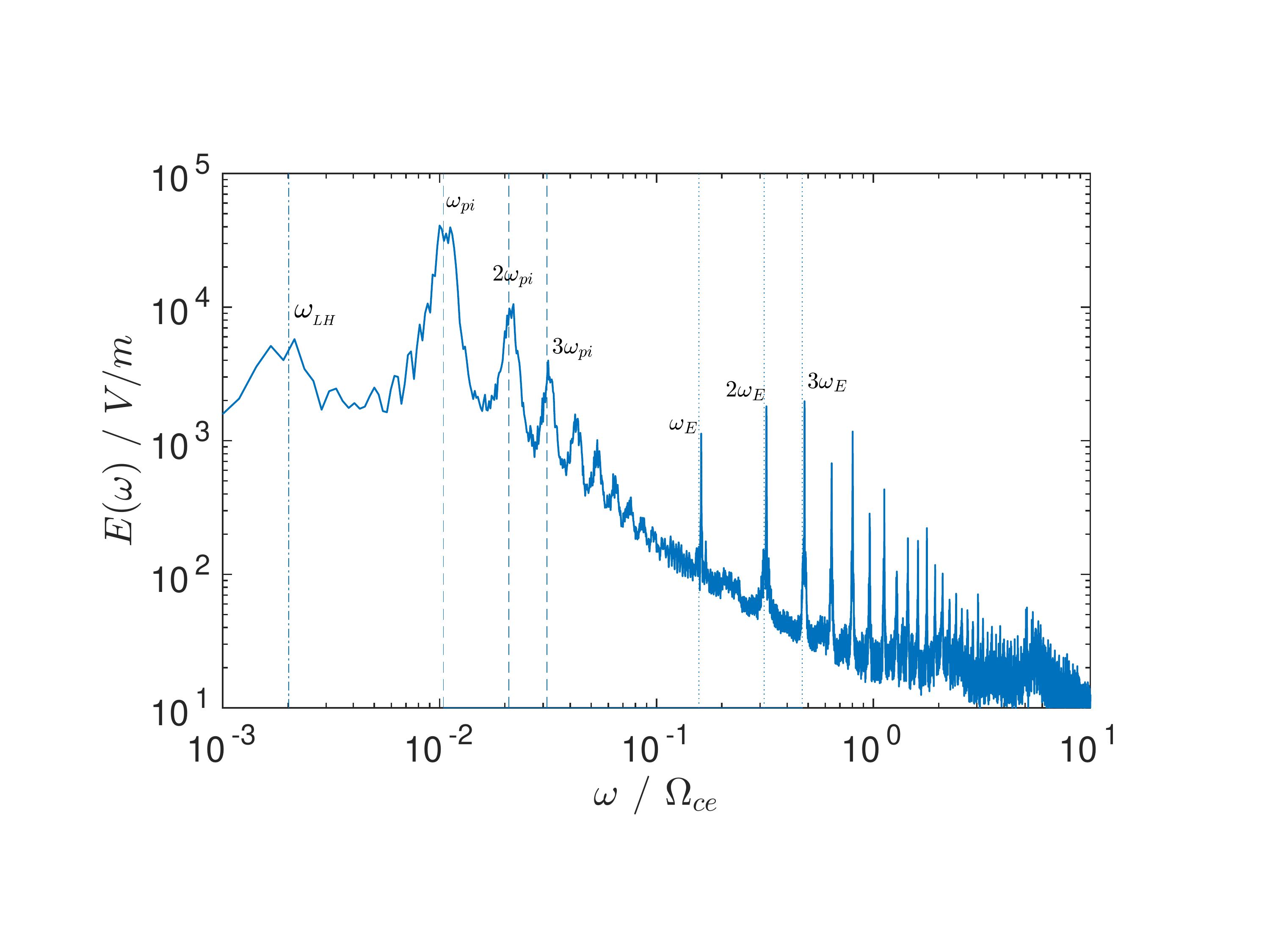}
  \caption{Maximal amplitudes for the $\omega$-spectrum for $E_x$, $\omega_E=V_e 2\pi /L$.}  
  \label{fig:enetempdist-c}
\end{figure}

In the nonlinear regime the perturbed electric field develops into a
robust quasi-coherent mode with a primary wave vector around the fundamental
cyclotron resonance.\cite{BoeufFP2014} The growing mode is driven by the
energy input from the few lowest order cyclotron resonances with $k_{0}$
providing the dominant contribution, as seen in Fig.~\ref{fig:density-d}.
The electron density, Fig.~\ref{fig:density-a}, is modulated mostly at
$k=k_{0}$. The ion dynamics has more complex structure showing nonlinearly
generated high-$k$ modes and ion trapping (bunching) features typical for
large amplitude nonlinear waves.\cite{Davidson_Nonlinear_methods} Features
of the localized ion trapping are further seen in the comparison of the
electric field and ion energy profiles, Fig. \ref{fig:density-c}, as
well as in the electric field structures correlated with the ion energy
fluctuations, Fig.~\ref{fig:density-c}. A characteristic nonlinear cnoidal
wave structure of the ion density, Fig.~\ref{fig:density-a}, is
further confirmed by the equal spacing of peaks in the $k$ and $\omega$
spectra in figures \ref{fig:density-c} and \ref{fig:enetempdist-c},
respectively.

\begin{figure}[thp]
   \includegraphics[width=0.975\columnwidth,viewport=77 74 665 507,clip]{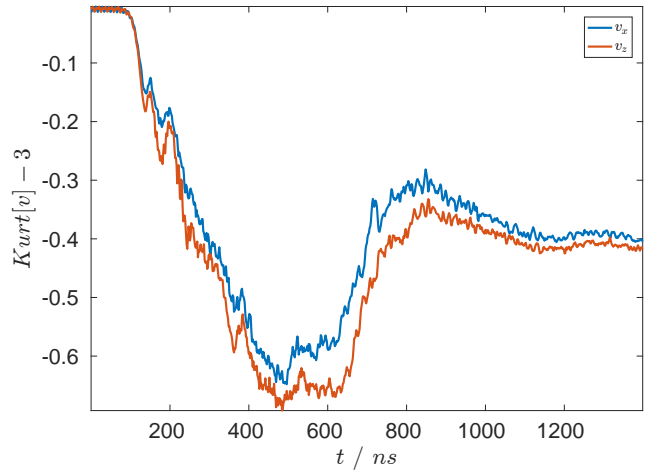} 
  \caption{Excess kurtosis ($4^{\text{th}}$ central moment) for the electron distribution over time. Note the slight asymmetry for $v_x$ and $v_z$.}
  \label{fig:enetempdist-d}
\end{figure}

The fluctuation spectra remain well quantized and predominantly retain
the fundamental cyclotron mode structure. However, one can also see
the development of a long wavelength envelope in the density
fluctuations analogous to the typical picture of the modulational
instability which is another evidence of the inverse cascade. The
modulations and cnoidal features increase in time, as shown in figures
\ref{fig:density-b} and \ref{fig:density-d}.

\section{Demagnetization of the electron motion and transition to the ion sound modes}

Many studies of the ECDI instability have emphasized the effects of
demagnetization of electrons and the transition of the mode into the regime
of the ion-sound instability that occurs in the absence of a magnetic field.
It is important to note that the mode structure and demagnetization
mechanism is a sensitive function \cite{ArefevTechPhys1970,LashmoreNF1973,LampePRL1971} of
the $k_{y}v_{e}$ parameter, where $k_{y}$ is the wave vector along the
magnetic field. Here we consider the case of strictly perpendicular
propagation, $k_{y}=0$.

The demagnetization of electron dynamics for large values 
$k^{2}\rho_{e}^{2}$ can be easily seen from equation (\ref{eq:dispersion-magn}).
In the limit $k^{2}\rho _{e}^{2}\gg 1$, the contribution of the terms $\exp
\left( -k^{2}\rho _{e}^{2}\right) I_{m}\left( k^{2}\rho _{e}^{2}\right)
\rightarrow 1/\left( k\rho _{e}\right) $ for all $m=0,1,...$ is neglected
and the electron response in Eq.~(\ref{eq:dispersion-magn}) becomes
$K_{e}=\left( k^{2}\lambda_{De}^{2}\right) ^{-1}$, which corresponds to
the Boltzmann response of unmagnetized electrons. 

The electron demagnetization in the linear short wavelength regime
$k\rho_{e} \gg 1$ can be viewed as the transition of the  lower-hybrid
mode (propagating strictly perpendicular to the magnetic field) to the ``high-frequency ion-sound''. Indeed, the dispersion relation for the
quasi-neutral lower hybrid mode with warm electrons has the form \cite{SmolyakovPPCF2017}
$\omega ^{2}=\omega _{LH}^{2}\left( 1+k_{\bot }^{2}\rho _{e}^{2}\right)$.
It is easy to see that in the short wavelength regime with
$k_{\bot}^{2}\rho _{e}^{2}\geq 1$, the mode dispersion relation becomes
$\omega^{2}=\omega _{LH}^{2}k_{\bot}^{2}\rho_{e}^{2}=k_{\bot}^{2}c_{s}^{2}$.
The lower hybrid mode is present in the measured frequency spectrum, as
shown in Fig.~\ref{fig:enetempdist-c}.

The neglect of all $m=1,2,3.. $ cyclotron harmonics in Eq. (\ref{eq:dispersion-magn}) may be justified for large $k\rho _{e} \gg 1 $, but not near the cyclotron resonances,
$\left[\left(\omega-k_{x}v_{E}\right) ^{2}-m^{2}\Omega_{ce}^{2}\right]\rightarrow 0$,
where these terms cannot be neglected. Therefore the mode properties may be
close to the lower-hybrid/ion sound mode which is determined by the first two terms in Eq.~(\ref{eq:dispersion-magn}), but the mode drive is
determined by the resonance $\left[\left(\omega-k_{x}v_{E}\right)^{2}-m^{2}\Omega_{ce}^{2}\right] \rightarrow 0$, where 
the $m=1$ is the most important. This resonance condition fixes the wavelength of the coherent mode at $k_0=\Omega_{ce}/v_{E}$. Note that in 
simulations the measured  phase velocity of the coherent wave turns out to be of the order of the ion sound
velocity within a factor of 2.

Deviations from quasi-neutrality bring in the effects of the electron
Debye length (or, Debye shielding), similar to the ion sound modes in
the short wavelength regime
$\omega^{2}=k^{2}c_{s}^{2}/(1+k^{2}\lambda_{D}^{2})$ so that $\omega
\rightarrow \omega_{pi}$ for $k^{2}\lambda_{D}^{2}>1$. The short
wavelength structures in the ion density are seen in the sharp peaks
of ion density which contain high $k$ modes (Figs.~\ref{fig:density-a}
and \ref{fig:density-b}) and explain the $\omega _{pi}$ (and its
harmonics) peaks in the amplitude spectrum, as shown in
Fig.~\ref{fig:enetempdist-c}.

\begin{figure}[thp]
  \includegraphics[width=0.975\columnwidth,viewport=30 69 751 523,clip]{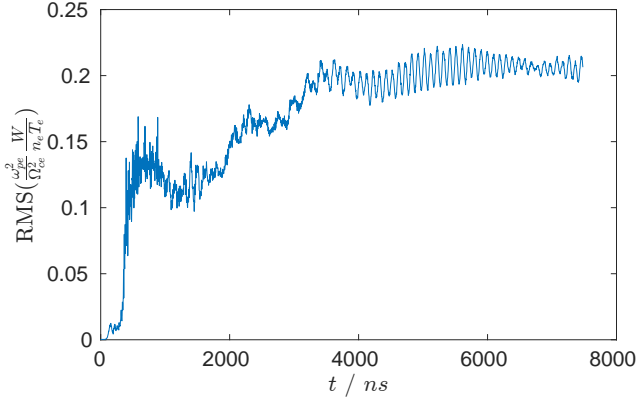}
  \caption{Parameter $\Xi$ in the condition for turbulent destruction of cyclotron resonances: $\Xi >\left( k\rho _{e}\right) ^{-1}$.}
  \label{fig:current-combi-d}
\end{figure}

\begin{figure}[thp]
  \includegraphics[width=0.975\columnwidth,viewport=74 63 781 564,clip]{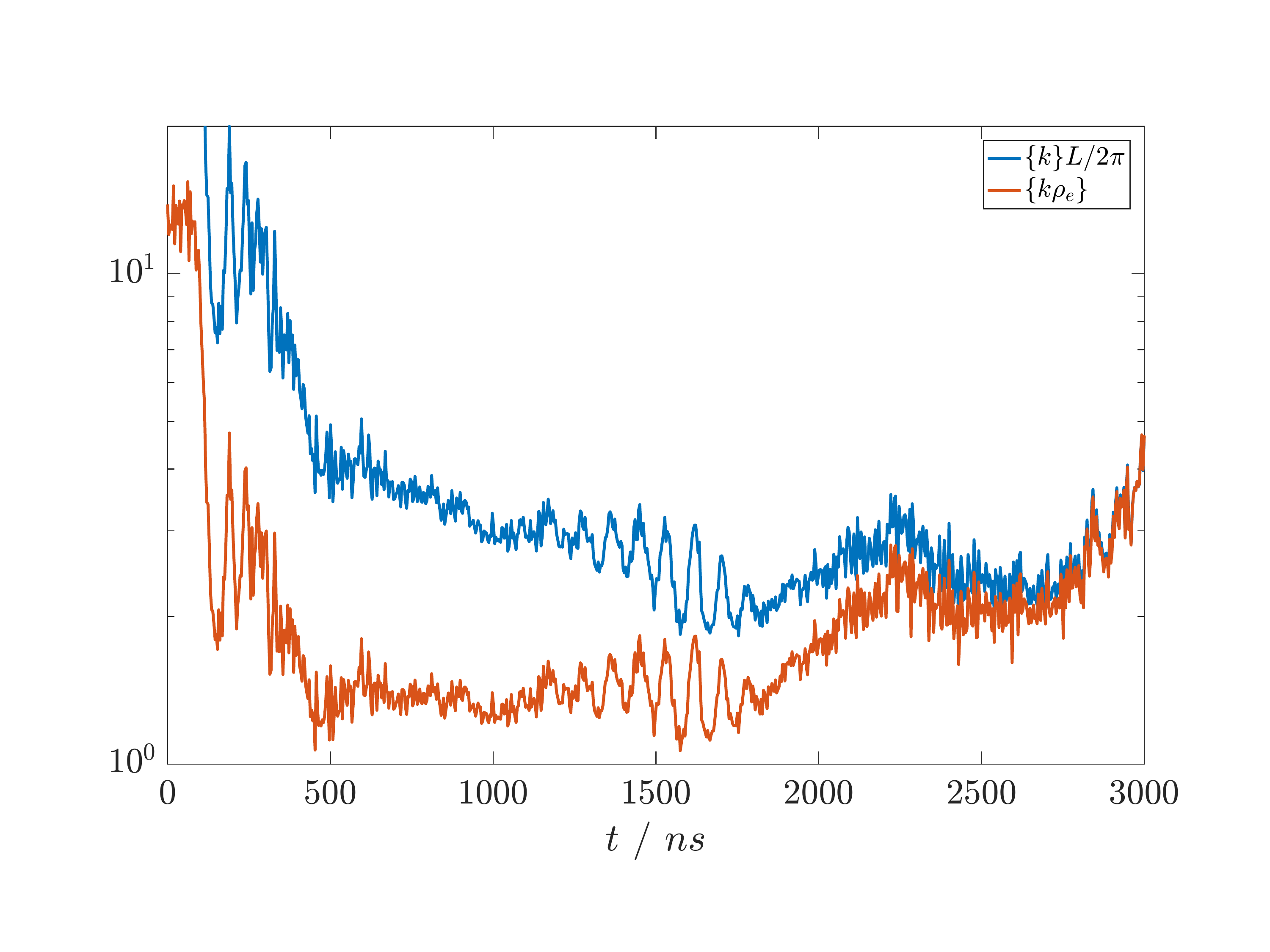} 
  \caption{Energy weighted averages of $k$ and $k\protect\rho _{e}$ from
anomalous current over time. See text for definition.}
  \label{fig:current-combi-b}
\end{figure}
%
%    time(us)  k_0 \rho_e
%    0.3288    1.5473
%    1.2203    3.0810
%    2.1081    4.8375
%    2.9959    6.4103

\begin{figure}[thp]
  \includegraphics[width=0.975\columnwidth,viewport=102 43 760 564,clip]{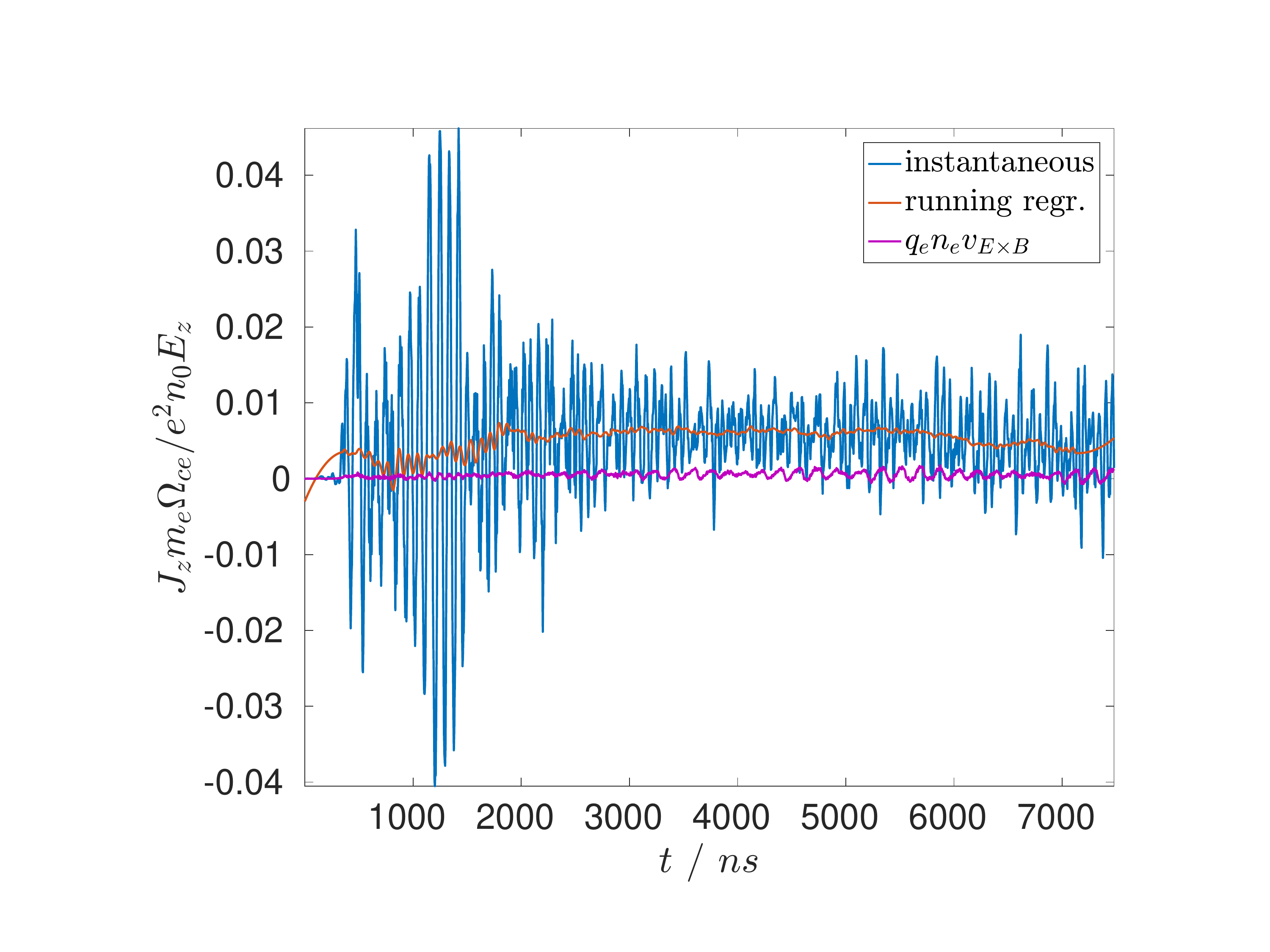} 
  \caption{Mean anomalous current density over time, in\-stan\-taneous value and a 2nd order running regression using a window much larger than the oscillation frequency, shown with the concurrent $E\times B$ flux. The latter is negligible.}
  \label{fig:current-combi-c}
\end{figure}

\begin{figure}[thp]
\includegraphics[width=0.975\columnwidth,viewport=70 94 742 533,clip]{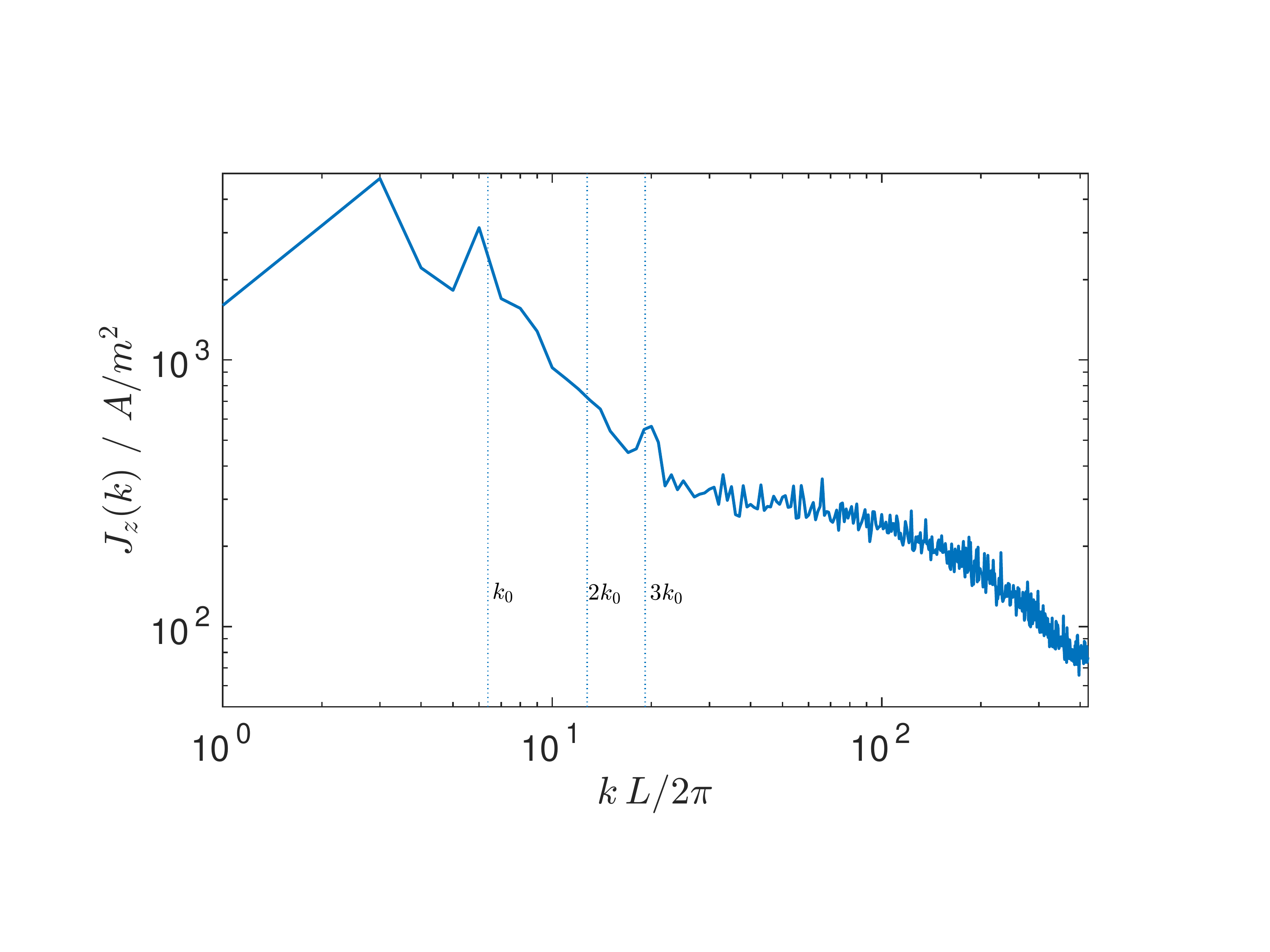}
\caption{Maximal amplitudes of the anomalous current $J_{z}$ as a
  function of the wave vector $k$ over the simulation. The
  sub-resonant (lower than $k_{0}=\Omega _{ce}/v_{E}$) components are
  observed to dominate over linearly resonant values. Note the finite
  value at system size.}
  \label{fig:current-combi-a}
\end{figure}

The cyclotron resonances can be destroyed by collisions even for
$\nu/\Omega _{ce}<1$ \cite{Pitaevskii1969,HubaPF1979}. The collisions
destroy the resonances when the particle diffuses by the distance
$\lambda/2=\pi/k$ over the period of the cyclotron rotation
$\tau _{c}=2\pi /\omega _{c}$,
or when $\delta R=\left( D_{c}\tau _{c}\right) ^{1/2}>\lambda /2$.
For the collisional diffusion with $\nu <\omega _{c}$, $D_{c}=\nu \rho
_{e}^{2}$, thus the collisions will destroy cyclotron resonances for
$\left(\nu /\Omega _{ce}\right) k^{2}\rho _{e}^{2}>\pi /2$.

A number of previous works have argued that nonlinear effects can also
effectively demagnetize the electrons via the anomalous resonance broadening
\cite{DumPRL1969}. A simple criterion for this may be obtained as follows.
Let us consider short wavelength modes with $k\rho_{e}\gg 1$. In this
regime, the electron experiences $N=2\pi \rho _{e}/\lambda $ scattering
events or ``collisions'' during one period of the cyclotron rotation. Each
``collision'' represents a small angle scattering with velocity change
$\delta v$: $m_{e}v_{e}\delta {v}=e\delta {\phi }$. During
such a "collision" the electron guiding center is shifted by the distance:
$\delta r=\delta v/\Omega _{ce}$. Each ``collision'' is random and the net
displacement $R$ over the time $\tau _{c}=2\pi /\Omega _{ce}$ is $R=\delta
rN^{1/2}$, giving the effective nonlinear diffusion coefficient
$D_{nl}=R^{2}/\tau _{c}=\Xi v_{Te}\lambda /4$, where $\Xi \equiv \left(
\omega _{pe}^{2}/\Omega _{ce}^{2}\right) W/\left( n_{0}T_{e}\right) $,
$W=E^{2}/8\pi $, where we have used $\delta \phi =E\lambda /2$,
$v_{e}=v_{Te}\equiv \left( 2T_{e}/m_{e}\right) ^{1/2}$. The cyclotron
resonances will be destroyed when over one cyclotron period $\tau _{c}$ the
particle is displaced due to nonlinear diffusion by a distance larger than
the half-wavelength, $\left( D_{nl}\tau _{c}\right) ^{1/2}>\lambda /2$. This
gives the criterion of nonlinear destruction of cyclotron
resonances as $\Xi >\left( k\rho _{e}\right) ^{-1}$.\cite{BiskampPF1973,LampePF1972a}

Alternatively, the effects of nonlinear resonance broadening can be
described by the addition of the nonlinear diffusion term $ik^{2}D_{nl}$
into $K_{e}$ in Eq.~(\ref{eq:dispersion-magn}). For large $k^{2}\rho
_{e}^{2}$, $\exp \left( -k^{2}\rho _{e}^{2}\right) I_{m}\left( k^{2}\rho
_{e}^{2}\right) =1/\left( k\rho _{e}\right) $, the summation of all
cyclotron harmonics can be performed giving \cite{LampePRL1971}
\begin{eqnarray}
K_{e}^{nl} &=&\frac{1}{k^{2}\lambda _{De}^{2}}\left[ 1+\left( \frac{\pi }{2}%
\right) ^{1/2}\frac{\left( \omega -kv_{E}\right) }{kv_{e}}\right. \times
\notag \\
&&\times \cot \left. \left( \pi \frac{\omega -kv_{E}+ik^{2}D_{nl}}{\Omega
_{ce}}\right) \right] .  \label{knl}
\end{eqnarray}%
For $k^{2}D_{nl}>\Omega _{ce}$, which is equivalent to the condition
$\left(D_{nl}\tau _{c}\right) ^{1/2}>\lambda /2)$,
$\cot \left(ik^{2}D_{nl}/\Omega _{ce}\right) \simeq -i$ and the equation
(\ref{knl}) corresponds to the response of unmagnetized electrons.

The destruction of cyclotron resonances was considered in
Ref.~\onlinecite{LampePRL1971} as the main nonlinear effect resulting in
saturation of electron cyclotron instability and transition to the
regime of slower ion sound instability in absence of the magnetic
field. In the course of the nonlinear evolution of the instability the
wave and electron thermal energy grow simultaneously,
Fig.~\ref{fig:enetempdist-a}. As a result, the parameter $\Xi$ remains
well under unity so that the condition $\Xi >\left( k\rho_{e}\right)
^{-1}$ is typically not satisfied,
Fig.~\ref{fig:current-combi-d}. Note that the effective $k\rho_{e}$
in our simulations remains of the order of unity,
Fig.~\ref{fig:current-combi-b}. The persistence of cyclotron
resonances is also evident in the spectrum, Fig.~\ref{fig:density-c},
which shows the frequency peaks at $k v_{E}=m\Omega _{ce}$.

Numerical noise may influence the results of particle-in-cell
simulations \cite{LangdonPF1979} by imitating the effects of
collisions. One can estimate the noise level by using the
fluctuation-dissipation theorem and assuming Poisson statistics for
electron and ion fluctuations. This yields an estimate for noise
energy $W_{noise}\approx n_{0}T_{0}/\sqrt{N_{p}}k\lambda_{D}$, where
$N_{p}$ is the number of particles within the wavelength $2\pi
/k$. Immediately it becomes clear that while high-$k$ modes may be
(ideally) well resolved, numerical noise is less efficiently damped by
plasma response in the low-$k$ region (which benefits from more
particles). We may therefore estimate the noise level as
$W_{noise}=T_{0}/\sqrt{2\pi
  N_{\lambda}N_{p}/(k_{L}\lambda_{D})}k_{L}\lambda _{D}$,
$k_{L}=2\pi/L$, which for our parameters gives us $\Xi =0.1$ with
$10^{4}$ particles/cell, and $\Xi=1$ for $10^{2}$ particles/cell using
$N_{\lambda }=8$. Therefore, electron demagnetization in part might be
attributed to particle noise in simulations where a low number of
particles is used, and certainly it may be argued that results from
such simulations will be noise-dominated. This is evident from
Fig.~\ref{fig:current-combi-d}; fluctuation levels in well-resolved
simulations are observed to be much lower than the higher noise
estimate.

\section{Anomalous current}
The ECDI instability could be one of possible sources of the anomalous
electron current (leading to anomalous mobility) in the direction of
the applied electric field, which is observed in many experiments with
$\mathbf{E\times B}$ plasmas.\cite{BoeufJAP2017}  In 1D simulations
the total current can be directly calculated \cite{BoeufFP2014} from
the particle distribution function $\mathbf{\Gamma}=\int
\mathbf{v}fd^{3}v$. The diagnostic of the anomalous current in the
simulations is another source of important information on the electron
dynamics. Fluctuating electric field in the $x-$direction in general
leads to particle displacement in the $z-$direction and thus may
contribute to the anomalous current $J_z=e\Gamma_z$.  Our simulations
show however that the anomalous current along the applied electric
field, $J_{z}$, is not due the $E\times B$ flux.  Figure
\ref{fig:current-combi-c} shows the instantaneous and running
(Savitzky-Golay\cite{orfanidis1996introduction}) average of $J_{z}$
current as well as the $\Gamma_{E\times B}=\left\langle
\widetilde{n}\widetilde{E}_{x}\right\rangle /B$ flux.  The
$\Gamma_{E\times B}$ flux is very small in our simulations as shown in
Fig.~\ref{fig:current-combi-c}, contrary to the results in
Ref.~\onlinecite{LafleurPoP2016a}.  Note that the current in the
direction of the $\mathbf{E\times B}$ drift is very close to the
current of magnetized electrons $\Gamma_{x}=nv_{Ex}$, where $n$ is the
total density and $v_{Ex}=-E_{0}/B$ is the equilibrium drift, as shown
in Fig.~\ref{fig:enetempdist-b}.

The large discrepancy of the total electron current ${\Gamma }_z=\int {v_z}fd^{3}v$ from the 
the $\widetilde{n}\widetilde{E}_{x}/B$ flux is not surprising for the electron-cyclotron drift modes.
The dominance of the
$E\times B$ flux (in $z$ direction) is expected only in the case of fully
magnetized electrons, and distinct time and length scale separation in the
electron velocity. The relation  $\Gamma _{z}\equiv \left\langle\int v_{z}fd^{3}v\right\rangle\simeq
\left\langle \widetilde{n}\widetilde{E}_{x}\right\rangle /B$ is only valid
when $v_{e}\simeq \widetilde{v}_{E}\gg \left( v_{I},v_{\pi }\right) $ where
the $\left( v_{I},v_{\pi }\right) $ are the inertial and viscous
contributions to the electron velocity \cite{SmolyakovPPCF2017} which are
small only for
$\omega \ll \Omega _{ce},k\rho _{e}\ll 1$ and $kv_{E0}\ll\Omega _{ce}$.
The latter condition is not satisfied for the cyclotron
resonance modes so that the electron velocity in $z$ direction deviates
significantly from $\widetilde{E}_{x}/B$: though the mode frequency is low
in the laboratory reference frame, $\omega \ll \Omega _{ce}$, the electrons
experience fast oscillating electric field due to the fast $E\times B$
motion when $kv_{E0}\simeq\Omega _{ce}$.

It is also worth noting that fully demagnetized electrons in the
ion-sound regime, like in Eq.~(\ref{knl}), which are not affected by the magnetic field, would not experience the $E\times B$ drift, and no anomalous
current in $z$ direction should be expected in this case. Therefore calculation of the anomalous electron current via the relation  $\widetilde{n}\widetilde{E}_{x}/B$ as in Ref.~\onlinecite{LafleurPoP2016b} is not justified for the fully demagnetized ion sound regime.    

Parameterizing the anomalous current in the form
$\Gamma_{z}=\left(\nu/\Omega _{ce}\right) _{eff}nE_{z}/B$ and noting that
$\Gamma _{x}=nE_{z}/B$ one can express the effective Hall parameter as
$\left( \nu /\Omega _{ce}\right) _{eff}=\Gamma _{z}/\Gamma _{x}$. In our
simulations we have $\left( \Omega _{ce}/\nu \right) _{eff}=165\pm 12$.
The values of $\Gamma_{z}$ and $\Gamma _{x}$ are also shown in the shift
of the center of the distribution function in  Fig.~\ref{fig:enetempdist-b}.

The spectrum of the anomalous current in $z$ direction, $J_{z}=e\int v_{z}fd^{3}v$ which also shows the presence of inverse cascade. 
As can be seen in Fig.~\ref{fig:current-combi-a}, low-$k$ modes are the
most effective in driving anomalous current, making the anomalous current
sensitive to the simulation box size. In the nonlinear stage, the
current peaks at the wavelengths well below of the lowest cyclotron
resonance mode $k_{0}$. Temporal evolution of the effective wave number is
illustrated in Fig.~\ref{fig:current-combi-b}, where we show the
characteristic $k$-value weighted with the squared $J_{z}$ current
amplitude: $\{k\}=\sum_{k}|J_{k}|^{2}k/\sum_{k}|J_{k}|^{2}$.  The
latter quantity can be viewed as an effective wave vector for the
``current center of mass'', using the energy of each mode as the weight.
To reduce the noise contribution, we impose a signal-to-noise ratio of 50
by thresholding (consistent with the 1\% noise estimate given above).
The anomalous current is dominated by the contribution from the
wavelength in the range $k\rho _{e}=1\div 2$.
In figure~\ref{fig:current-combi-a} we also show the weighted average
for $k\rho_e$.

\section{Effect of energy losses}

\begin{figure}[thp]
  \includegraphics[width=0.975\columnwidth,viewport=46 119 744 505,clip]{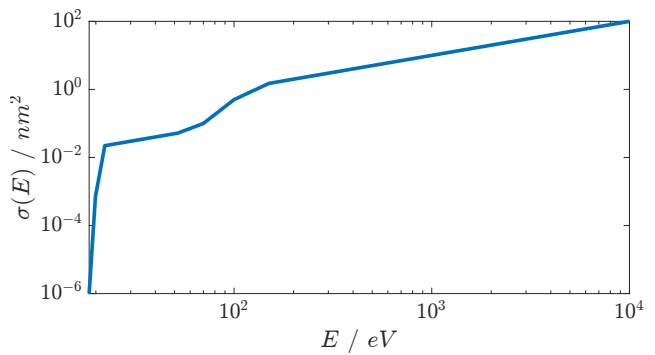}
  \caption{Cross section for MC collisions for parallel energy loss modeling. Threshold energy is $17.5\,$eV, which is subtracted from electron energy in the event of collision.}
  \label{fig:cross-sect}
\end{figure}

Nonlinear simulations demonstrate that ECDI is a very effective
mechanism of electron heating.\cite{AdamPoP2004} In our simulations,
even when started from the low energy of a few eV, over the simulation
time of a few ${\mu}$s, the electron energy grows to 100s eV, which are
unrealistically large values of electron temperature for a Hall-effect
thruster plasma. There are several loss mechanisms that are operative
in experimental settings. One such mechanism is parallel (to the
magnetic field) losses of high-energy electrons into the sheath
loss-cone, when a finite length of plasma along the magnetic field is
considered in spatially 2D and 3D simulations. Even though ECDI
heating occurs expressly in the perpendicular velocity components, we
may assume that the particles occasionally experience collisions, and
this way a high perpendicular energy will reflect to a high parallel
velocity that incurs parallel losses. A lower energy particle then moves
into the region to avoid a loss of total number of particles (fast
parallel transport). This process may be viewed as an excitation
collision scattering with a background
plasma\cite{KaganovichPOP2007,SydorenkoTh2006}, using the
cross-section shown in Fig.~\ref{fig:cross-sect}). In a Monte Carlo
sense this process utilizes the null-collision model, where the
collision probability for a particle of a certain energy within a time
step $\Delta\hspace{-0.25ex}t$ is
$P=1-\exp{(-\Delta\hspace{-0.25ex}t\,\nu(E))}$ where $\nu(E)=v_j
\sigma(E) n_a$. Here $v_j$ is the particle velocity, $\sigma(E)$ is
the collisional cross section for the particle energy $E$, and $n_a$
density of the background (only used for this purpose, and here chosen
to be $3\cdot10^{19}\,m^{-3}$). In the event of a collision, the
threshold energy is subtracted from the particle energy, and the velocity
components are modified with the scattering Euler angles.

\begin{figure}[thp]
  \includegraphics[width=0.975\columnwidth,viewport=53 65 769 573,clip]{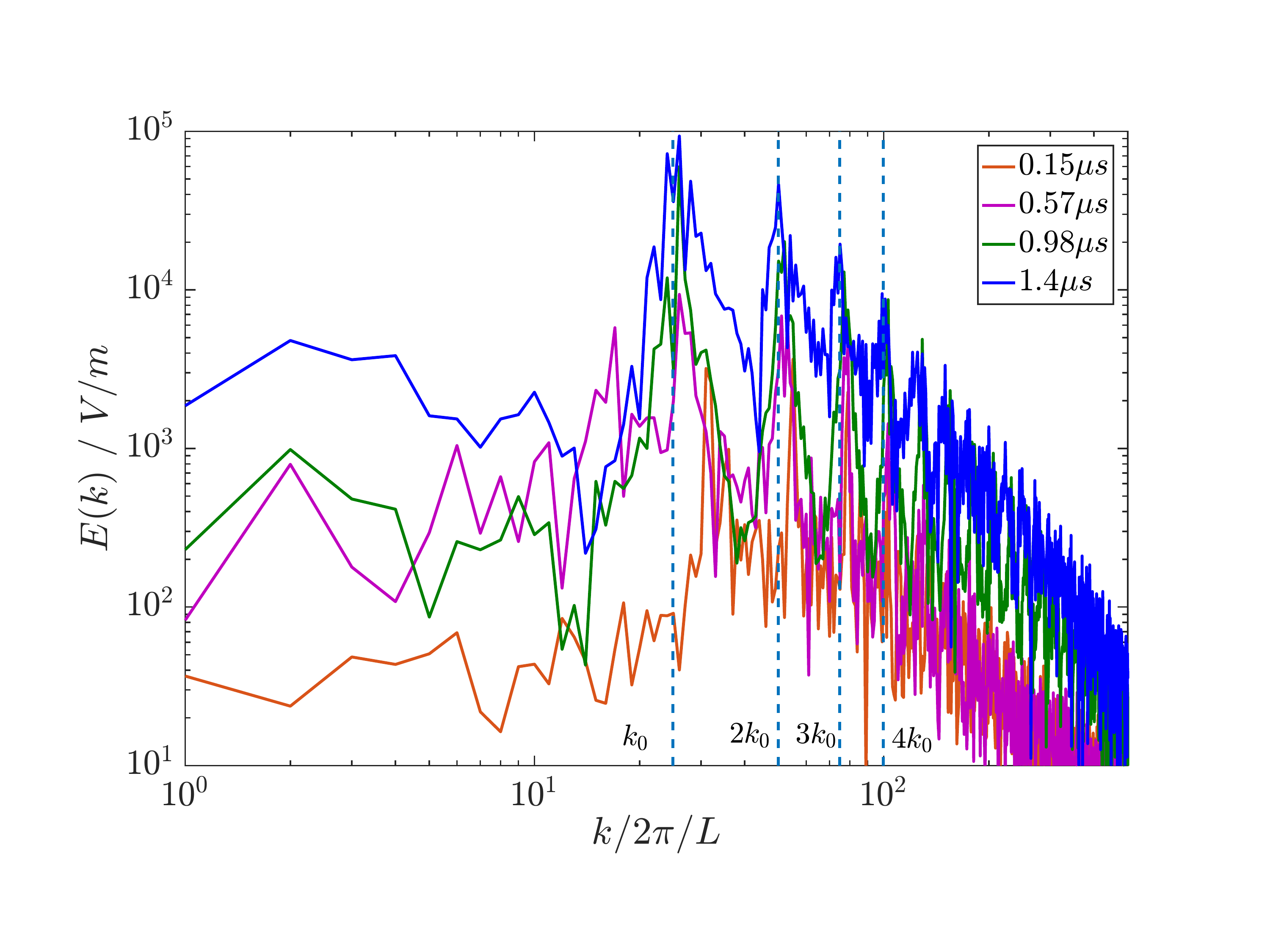}
  \caption{Fluctuation wave number spectrum for the case with electron energy losses.}
  \label{fig:density-spec}
\end{figure}

\begin{figure}[thp]
  \includegraphics[width=0.975\columnwidth,viewport=45 26 769 581,clip]{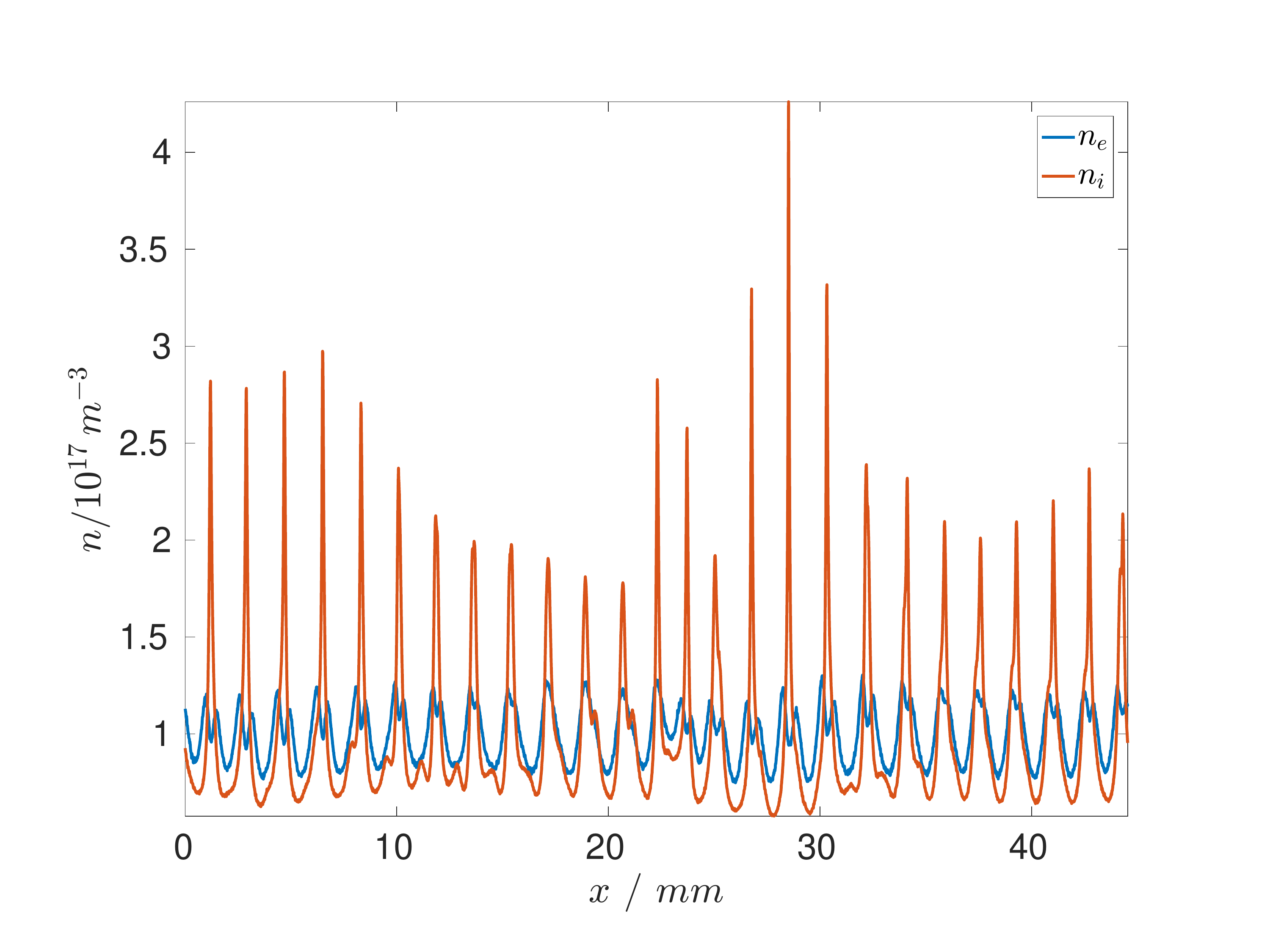}
  \caption{Ion and electron densities when energy loss is introduced for electrons.  Ion density fluctuations are increased, compare with figure~\ref{fig:density-a}.}
  \label{fig:density-end}
\end{figure}

We show in figure \ref{fig:density-spec} that the electric field
spectrum in nonlinear regime remains largely unaffected as compared to
the case without losses. It is interesting that losses increase ion
density fluctuations quite significantly, making the density
fluctuations even more peaked, as shown in figure
\ref{fig:density-end}. Based on these results we expect the cyclotron
resonances in plasmas with parallel losses to be even more strongly pronounced because the destruction of resonances is more effective for
higher electron temperature. Also, the linear drive remains effective
because the electrons are continually being re-circulated into the
vicinity of the cyclotron resonance. Therefore, parallel losses are
unlikely to modify the nonlinear features.

\begin{figure}[thp]
  \includegraphics[width=0.975\columnwidth,viewport=43 96 706 489,clip]{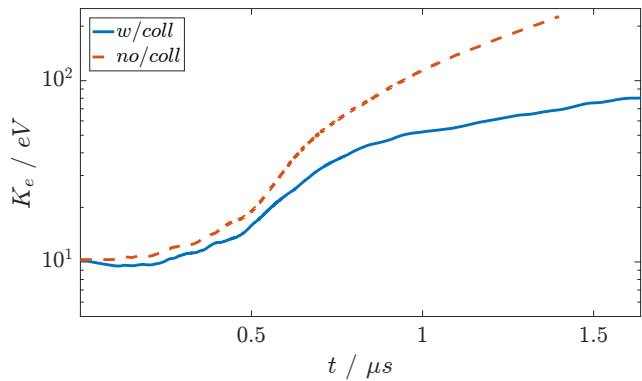}
  \caption{Temperature evolution of electrons with and without collisions.}
  \label{fig:heating-coll}
\end{figure}

\section{Summary}

We have investigated the dynamics of electron cyclotron drift
instability using highly resolved particle-in-cell simulations in 1D3V
with realistic mass ratios and using parameters relevant to the
Hall-effect thruster. The large simulation box allowed for
investigations of large scale nonlinear dynamics of ECDI pumped by
transverse $\mathbf{E\times B}$ current. In the nonlinear regime
we observe a large amplitude coherent mode (periodic cnoidal wave)
driven mainly at the electron cyclotron drift cyclotron resonance
$k_0=\Omega_{ce}/v_E$. High $k$ mode generation occurs due to wave
focusing (sharpening) associated with nonlinear ion breaking
\cite{Davidson_Nonlinear_methods}, particularly evident in the ion
density fluctuations. Simultaneously, we observe energy flow to long wave
length and low frequency modes manifested by the generation of the long
wavelength envelope. The long wavelength oscillations in our simulations
develop on the $\mu$s time scale (or a little faster)  and these modes could
be similar to  the low frequency features that were found in recent
experimental observation of the $\mathbf{E\times B}$ instability
\cite{TsikataPRL2015}. The long wavelength modulations in our
simulations also resemble some nonlinear features of the electron
cyclotron modes observed in Earth's bow shock\cite{BrenemanJGR2013}.
The generation of the long wavelength mode and associated with it energy
transfer to the long wavelength part of the spectrum is expected to be
important in the mode saturation mechanism along with possible parametric
instabilities of large amplitude
waves.\cite{AkhiezerPhysLettA1998,MikhailenkopoP2003}

We have shown here that demagnetization criterion due to nonlinear
resonance broadening (and overlapping) is not fulfilled for electrons in
our simulations. The electron cyclotron resonances remain prominently
evident, especially at low $m$, while higher resonances become
sub-dominant, which is similar to the results of other simulations of
electron-cyclotron instability performed for space conditions
\cite{MuschiettiJGR2013,MuschiettiAdvSpR2006}.
The full demagnetization of the electron response requires two conditions:
$k \rho_e \gg 1$ -- the modes have to be in the short wavelength regime,
and $\Xi  k \rho_e  >1$  -- for the nonlinear destruction of the cyclotron
resonances.
These two conditions (formally equivalent to the limit of zero magnetic
field, $B \rightarrow 0))$ result in the fully demagnetized electron
response and the resonant drive fully equivalent to that of the beam of
unmagnetized electrons. For turbulent fluctuations in our simulations, the
condition $k \rho_e \gg 1$, is only marginally exceeded, see
Fig.~\ref{fig:current-combi-b}, while the  condition  $\Xi  k \rho_e  >1$
is not satisfied, see Fig.~\ref{fig:current-combi-d}.  This suggests that
the magnetic field remains important in the mechanism of the instability,
electron heating and transport. 

Our simulations show that overall electron dynamics is dominated by the
cascade to long wavelength, low frequency modes down to the lower hybrid
range and below.  An important conclusion from our simulations is that the anomalous
electron current is dominated by the contributions from long wavelength
(sub-cyclotron-resonance harmonics) modes, from a few mm up to the box
size, Fig.~\ref{fig:current-combi-a}. This feature  is consistent with
experimental observations in which a significant fraction of the anomalous
current is directed through the large scale spoke structure  \cite{ParkerAppPhysL2010}. 
%
% Added due to prompting by the reviewer:
We speculate that while the energy input via the resonant ECDI may occur at small scales, the  nonlinear inverse cascade analogous to our 1D case result in energy condensation in large-scale structures, as also shown by analytical theory in Ref. \onlinecite{Lakhin}. 

Our simulations, while demonstrating important features of the
electron cyclotron modes driven by $\mathbf{E \times B}$ current, have
certain limitations due to their 1D nature. In general, the fluctuations
are expected to have 3D structure as experimental measurements indicate
\cite{TsikataPOP2010}. There are several ways in which fluctuations and
transport in general 3D case may differ from a simple 1D case.

First, when both components of the fluctuating electric field in the
plane perpendicular to the magnetic field are present, one can expect
that anomalous contributions both in $\mathbf{E}$ and $\mathbf{E\times B}$
directions will be present (see also the discussion in Section V above) and
thus modify the total current in the direction of the applied electric
field. The external electric field will have to be determined
self-consistently \cite{BoeufJAP2017} as result of the balance of
fluctuation energy (and the resulting anomalous current) and the externally
applied potential difference. 

Another important point is that fluctuations with a finite value of the
wave vector along the magnetic field, $k_\parallel$, may have different
dispersion properties and instability conditions.  As it was shown in
Refs.~\onlinecite{GaryJPP1970a,GaryJPP1971a,ArefevTechPhys1970}, and also
more recently in Ref.~\onlinecite{CavalierPoP2013}, the short wavelength
instabilities with significantly large values of
$k_\parallel \rho_e\geq O(1)$ reduce to the unmagnetized (ion-sound) form.
The actual 3D structure of instable modes and its role in the linear and
nonlinear development of unstable modes has to be determined in
self-consistent simulations resolving the direction along the magnetic field
\cite{CroesPSST2017} and proper account of sheath boundary conditions
\cite{SmolyakovPRL2013}. In our simulations, only an approximate
model of parallel losses was used to limit the saturation amplitude
for unstable modes. Saturation mechanisms that ultimately will define the
mode amplitude are sensitive to the particle and energy losses
\cite{BoeufFP2014,BoeufJAP2017} including those along the magnetic
field as well as ionization effects which are also important for
$E\times B $ plasmas\cite{EscobarPoP2014,EscobarPoP2015}. Comprehensive
account of all these effects also has to be done in full cylindrical
geometry\cite{CarlssonIEPC2015}. 
However, even in 1D simulations the importance of good resolution and a sufficiently large simulation domain
becomes apparent.

In general 2D and 3D cases, the gradient-driven and lower hybrid type
instability will be operative
\cite{DavidsonNF1975,SmolyakovPPCF2017,MatsukiyoJGR2009}.
%
% Added due to prompting by the reviewer:
One can therefore expect that the energy accumulation in long-wavelength
modes and contribution to the anomalous transport will be further enhanced
by the gradient-drift instabilities which generally have longer wavelengths
compared to the cyclotron modes studied here and will be directly active
in the mesoscale part of spectrum; between the external scale (of the order of the
size of the device) and small scales of the unstable modes.  

Part of this picture is the excitation of the gradient driven modes inside large scale structures as seen in as well in PIC simulations that show the $\lambda=4\,$mm wavelength
fluctuations inside the spoke\cite{MatyashIEPC2013}. In our periodic simulations, the external length scale is limited by the simulation box size. In realistic 2D/3D simulations this size can be related to the geometric size, like the lowest $m=1$ mode for cylindrical geometry. Additional processes as energy losses to the wall and ionization will also affect the scale of the large sale structure and reduce the fluctuation amplitude. In our simulations the fluctuation amplitude is of the order of the equilibrium electric field, while the experimental values are much lower. \cite{TsikataPOP2010}.   

This work is supported in part by NSERC Canada, and US Air Force Office of
Scientific Research FA9550-15-1-0226. Compute/Calcul Canada computational
resources were used. We would like thank E. Startsev (PPPL) for fruitful discussions.

\section*{References}

\bibliography{MyReferences}
\end{document}